\documentclass[11pt]{article}
\begin{document}
\title{ON THE MAGNETISM OF STARS AND PLANETS}
\author{S. N. Arteha}
\date{Space Research Institute, Profsoyuznaya 84/ 32, Moscow 117810, Russia}
\maketitle

\begin{abstract}

The self-consistent statistical approach to the problem of
planetary and stellar magnetism is suggested. 
The mechanism of magnetic fields generation in the 
astronomical objects, where the existence of fields is associated with the 
axial rotation of objects, is discussed. In the general case
the light pressure, the centrifugal, gravitational and other
forces produce partial $\rho$-separation 
of the charges. As a result of the system rotation, the magnetic fields 
of the currents of these charges are not compensated. The influence of
various factors on the magnetic field of some object is
analysed. 
\end{abstract}

\section{Introduction}

The origin of magnetism in various astronomical objects is of great 
theoretical and practical interest. The existence of magnetic fields 
in these cases indicates that currents are circulating inside such 
objects. What is the nature of these currents?

At the present time there is a great deal of information on this 
subject. In the literature the existence of magnetic fields 
is associated with: 

1) various dynamo processes: a) the classical 
kinematic dynamo-models (Busse, 1979; Parker, 1979; Moffat, 1978); 
b) the kink instability mechanism (Alfven et al., 1974; Alfven, 1981);  
c) the action of gravitational forces (Hide, 1956); 
d) the specific models (heat convection, precession) (Urey, 1952; 
Busse, 1976; Bullard et al., 1971; Runcorn, 1975); 

2) or with the remanent magnetization (Sharpe et al., 1976).

These theories encounter some problems: (Runcorn, 1975; Malkus, 1963; 
Stevenson, 1974); the Earth's core paradox (Higgins et al., 1971; 
Kennedy et al., 1973); the energetic contradictions (Jacobs et al., 1972); 
the contradictions of precession mechanism (Rochester et al., 1975; 
Busse, 1971) etc.

A variety of approaches to the problem of magnetism of astronomical
objects may exist. One of them consists in
describing the process of magnetic field birth and generation
up to observable values. This approach is undoubtedly
important and interesting, from principal positions, to elucidate the 
mechanism of such a generation. As an example we refer to the "battery 
theory" for the generation of cosmic magnetic fields (Biermann 1950; 
Mestel et al., 1962; Dolginov, 1988). Note, that the paper is 
concerned with the above mentioned theory (in initial cause). However, 
some differences exist. The basis for the "battery theory" is the 
"Biermann effect", where the following term provides the magnetic field 
generation:
$$
c{\bf \nabla}\eta\times{\bf \nabla}T = -{c\over e}{\bf \nabla}\times
{{\bf \nabla}P_e\over N_e} = {1\over 2}M_i{\bf \nabla}\times\vec{g_{eff}}, 
$$
where $\eta$ is the thermopower constant, $P_e$ is the electron pressure, 
$\vec{g_{eff}}=\vec{g}+\vec{\Omega}\times\vec{\Omega}\times\vec{r}$ is 
the sum of gravitational and centrifugal accelerations (the causes of 
generation: ~nonuniform rotation, the inhomogeneous distribution of 
chemical elements, temperature inhomogeneities). 
At the same time, however, it has rather essential disadvantages.
First, one must proceed from some unknown hypothetic state.
Second, for the time during which the magnetic field is
increasing, so many various random events take place, that
they can not be taken into consideration even in principle. For example, 
random flows of charged particles (cosmic rays) can violate the 
$2^n$ magnification (as a consequence of long-range Coulomb interactions). 
Third, even the main factors can be taken into account within
the framework of approximate models only (the continuum medium
model, for instance). Besides, there exist problems for calculation. 
At present it is impossible to solve the full system of equations 
(Dolginov, 1988). The magnitude of the stationary field cannot be 
determined from the linearized equations (note that the ~nonlinear 
terms are the same order of magnitude as the linear ones). But investigation 
of nonlinear effects are still in their infancy. 
All these difficulties do not allow quantitative 
characteristics of the field to be determined. There
exists, however, some another approach to this problem. It is
as follows. At present, a set of observable characteristics
exists for studying astronomical objects. These
characteristics (seem to be stable) include such ones, as the frequency of
rotation of a system, its composition, electric and magnetic
fields, etc. Here, the question arises: is there any
connection between these observable quantities, so that,
proceeding from some particular observable system's
parameters, one can calculate other parameters. 

Many astronomical objects (nebulae, planets, 
the Sun and other stars) are known to be rotating bodies. The attempt to find 
the relation between these phenomena is made in this
article. Since some astronomical magnetic fields are rather
stable, it can be supposed stable or quasistable
state exists in the system and the microscopic statistical approach to this problem can be
proposed. Developing this approach, an attempt is made
in this paper to describe theoretically stable states of the
system and the dynamics of fields. In principle, the statistical approach 
allows the exact solution for the stationary field to be obtained in 
integral form. Some distinct value of the field (to sufficient accuracy) 
can be calculated from this solution for all specific values of 
parameters (~nonlinear effects are implicitly taken into consideration). 

The peculiarities of this work are as follows:

- the mechanism dispenses with the need for initial magnetic fields;

- the mechanism proposed can explain the existence of magnetic fields in 
a wide majority of astronomical objects, i.e. it is rather universal; 

- the mechanism has self-consistent characteristics and leans upon 
the microscopic properties of matter, rather than upon the continuous 
medium models; 

- the mechanism is simple enough, that is, there is no need 
to make any additional assumption on properties of the inner region 
of objects (good conduction, for example) or on processes occurring 
inside the objects (convection, for example).

Note, that the suggestion of the
statistical approach to the problem is the most important part
of the article, since there exist a possibility to study
magnetic fields without knowledge about the nature of electric
double layers: the statistical method can include different
mechanisms which can lead to separation of charges (but not
only suggested). For example, the reverse problem can be
formulated: knowing fields, to find the possible position of
electric double layers and theirs magnitudes. After that the
discussion of magnetism (concrete mechanism) is more
objective. However, it is not our intention to follow this way
in the article. 

In section 2.1 the separation of charges and the magnetic field appearance 
are investigated for the rotating plasma model. This model can be 
applied to some nebulae, for example. In section 2.2 the action of a 
gravitational force in the compact rotating system is additionally considered. 
This model can explain the existence of stellar magnetism. In section 2.3 
the magnetism of planets (some "cooled" objects) is discussed. In this case the 
"cooled" (magnetic) materials in the surface layer of a planet (the Earth 
crust, for example) at temperature less than the Curie temperature can 
influence the magnetic field. In section 2.4 the general
remarks on the Kauling theorem and different mechanisms, and
some evaluations are presented. Section 3.1 outlines
self-sustained rotation of a system and
the equilibrium configuration of electric and magnetic fields.
Section 3.2 analyses the influence of various internal and
external factors and the dynamics of fields. 

\section{Models}

\subsection{The magnetic field in rotating plasma}                       

The mechanism of magnetic field generation in the astronomical objects 
can be qualitatively explained in terms of this model. The rotating plasma 
model can be applied for explanation of magnetic fields existence in 
rotating plasmoids, such as ionized cloud, hot nebulae, etc. 

As a consequence of long-range interaction between the charged particles, the 
plasma essentially differs from gas in some respects. The rotating plasma has 
some peculiarities as compared to rotating objects involving neutral particles. 
The existence of the centrifugal force
\begin{equation}
{\bf F_c} = m\Omega^{2}\rho{\bf }e_{\rho},
\end{equation}
(where ${\bf \Omega}$ is the angular frequency of the system, $m$ is the particle mass, 
$\rho$ is its distance from the axis of rotation) causes 
different effects on particles of different masses. As a result of 
these different effects, the $\rho$-dependences of particle concentrations 
are bound to be different for particles of different masses. 

According to the Boltzmann distribution,
\begin{equation}
n_{\alpha} = n_{0\alpha}\exp{({-U_{\alpha}\over kT})},
\end{equation}
where $T$ is the system temperature, $k$ is the Boltzmann constant, 
$n_{0\alpha}$ is the particles concentration on the axis of rotation, $U_{\alpha}$ 
is the potential energy of particles of $\alpha$-sort; for neutral particles 
\begin{equation}
U_{\alpha} = -{m_{\alpha}\Omega^{2}\rho^{2}\over 2}.
\end{equation}
Since the particles of different masses 
in plasma have different charges, this partial separation of 
particles produces a partial separation of charges, that is, the 
negatively charged region must lie near the axis of rotation, whereas the 
positiveLy charged region must lie near the system boundary 
(the distance $R_{0}$ from the axis of rotation). The electric field ${\bf E}_{0}(\rho,z)$ 
(in the polar coordinate system) exists as a result of this separation of 
charges.

This field opposes the considerable separation of charges. Therefore, 
the effect of partial separation of charges is bound to be often negligible 
for small systems, whereas the separation of charges can exert 
considerable influence on some physical characteristics of astronomical 
objects. 

There is some distance from the axis of rotation 
$R_{1}(z)$, where the local charge equals zero (some transverse section to the 
axis). For $0 < \rho < R_{1}(z)$ the plasma is negatively charged 
on the average; for $R_{1}(z) < \rho \leq R_{0}$ the plasma is 
positively charged. Therefore, the field ${\bf E}_{0}(\rho,z)$ at $R_{1}(0)$ 
is directed towards the axis of rotation. 

Of mine interest here is the fact, that the partial separation of charges 
gives rise to the $\rho$-dependence of charge density, that is, the plasma 
possessing a given charge density moves round a circle of distinct radius. 
Therefore, the circulating currents are inside the rotating plasma despite 
the fact, that all particles revolve with ${\bf \Omega}$ on the average 
(${\bf \Omega}$ has a distinct direction). For $R_{1} < \rho
\leq R_{0}$ the current flows in the direction of rotation
(since the charge is positive), whereas for $0 < \rho < R_{1}$ 
the current is opposite to the direction of plasma rotation (since 
the charge is negative). 
In the general case the magnetic actions of these currents are not compensated 
and the magnetic field exists. This is the magnetic field of a solenoid whose 
axis coincides with the axis of rotation of the system. 

It follows from the Biot-Savart law, that the magnetic field is 
\begin{equation}
{\bf H} = {1\over c}\int_{(V)} {[{\bf j'} \times {\bf R'}]\over R'^{3}}dv',
\end{equation}
where $V$ is the volume of the system, $c$ is the speed of light, $dv'$ is the region with currents, $R'$ is the 
distance of this region to the point of observation, $\bf{j'}$ is the 
current density. We have 
$$
{\bf R}' = (R_{\rho};R_{\varphi};R_z) = (\rho -
\rho_1\cos(\varphi_1-\varphi); \rho_1\sin(\varphi-\varphi_1); z-z_1);
$$
$dv'= \rho_1d\rho_1d\varphi_1dz_1~, ~~~ {\bf j}' =
(0;j_{\varphi};0)~, ~~~ j_{\varphi}(\rho_1,\varphi_1,z_1)=
e\Omega\rho_1q(\rho_1,\varphi_1,z_1).$ \\
The electric charge density $q$ is 
\begin{equation} 
q(\rho_{1}, z_{1}) =  
n_{0i}\exp{\Biggl ( -{1\over kT}U_{i}\Biggr ) } - n_{0e}\exp{\Biggl ( -{1\over kT}U_{e}\Biggr ) }, 
\end{equation} 
For rotating, fully ionized plasma, which consists of the elements with 
atomic number $N_{i}$, it follows in polar coordinates $\rho, \varphi, z$  
(axis $z$ is in the ${\bf \Omega}$ direction), that 
$$
U_{e} = -{m\Omega^{2}\rho_{1}^{2}\over 2} + 
e\int_{0}^{\rho_{1}} E_{0\rho}(\rho',z_{1})d\rho' +
e\int_{0}^{z_{1}} E_{0z}(\rho_{1},z')dz' + 
$$
\begin{equation}
\int_{0}^{\rho_{1}} {eH_z\over c}\Omega\rho'd\rho' -
\int_{0}^{z_{1}} {eH_{\rho}\over c}\Omega\rho_1dz'~ ,
\end{equation}
$$     
U_{i} = -{M\Omega^{2}\rho_{1}^{2}\over 2} 
- eN_{i}\int_{0}^{\rho_{1}} 
E_{0\rho}(\rho',z_{1})d\rho' - eN_{i}\int_{0}^{z_{1}} E_{0z}(\rho_{1},z')dz' -
$$
\begin{equation}
\int_{0}^{\rho_{1}} {eH_z\over c}\Omega\rho'd\rho' +
\int_{0}^{z_{1}} {eH_{\rho}\over c}\Omega\rho_1dz'~ ,  
\end{equation}
where $n_{e}(\rho,z)$ is the concentration of electrons, $n_{i}(\rho,z)$ is the 
concentration of ions, $-e$ is the electron charge ($e>0$), $m$ is the 
electron mass, $M$ is the ion mass, the electron and ion temperatures 
$T_{e} = T_{i} = T$; the quantities $n_{0\alpha}$ can be found from conditions: ~ $\int_{(V)} n_{\alpha}dv' = N_{\alpha}K_{\alpha},$ where 
$K_{\alpha}$ is the total number of particles of
$\alpha$-sort, $N_e=1$ and $K_{e} = N_{i}K_{i} + \Delta$, $\Delta$ is
the electron surplus. 

The electric field can be found from the integral equation
$$ 
{\bf E}_{0}(\rho,z) = \int_{(V)} {e{\bf R}'(\rho',z')q(\rho',z')dv'\over R'^{3}}
$$

Heat movements counteract the compensation of the charge density: the 
statistical (dynamical) equilibrium is established.
In the general case for rotating plasma of complex composition, 
$j_{\varphi} \rightarrow \sum_{l} j_{\varphi}^{l}~,~~
n_{0\alpha} \rightarrow n_{0\alpha}^{(l)}$. In principle, we can introduce 
the "effective" ion mass $M_i^{eff}$ (in the general case $M_i^{eff}\geq 
M_i^{av}=(\sum M_iK_i)/(\sum K_i)$) and $N_i^{eff}$. Note, that a small 
quantity of heavy ions plays an important role. As follows from the derived 
expressions, the effect is proportional to the particle's concentration 
(or, the electron surplus $\Delta$). Since for $\Omega =0$ and for $\Omega
\rightarrow\infty$ the effect vanishes, there exists such a value $\Omega_m$ 
that $\mid H_z(\Omega_m)\mid = max$. Analogously, there exists $T_m$ for which 
$\mid H_z(T_m)\mid = max$. The dependence $H_z(\Omega,T)$ can be represented 
as a "pit". This is only the discussion of principles of approaches to the 
magnetic field generation problem. Strictly speaking, the model uses the 
externally given fixed "wall". To use the described model for real objects, the 
following factors need to be taken into consideration: the gravitational 
force, if it is important (maintains the system as a whole, for 
example), or, the expansion of system boundary (the change of integration 
limits, if the equilibrium distribution has a chance to be set) and the 
dependence $\Omega(\rho,\varphi,z)$ (nonrigid rotation). 
Some practical remarks will be given in Section 2.4. 

\subsection{The magnetic fields of rotating stars}

The consideration of the starry magnetism differs from the case discussed 
above both quantitatively (${\bf \Omega}$ and $n_{0\alpha}$ may be 
considerably larger than corresponding quantities for the ionized cloud) and 
qualitatively. One of such qualitative differences lies in the fact, that 
the gravitational force must be taken into account (the external boundary 
doesn't be fixed, but can be determined in the self-consistent manner): 
\begin{equation}
{\bf F_g} = \gamma{M(r)m{\bf r}\over r^{3}}~,
\end{equation}
where $\gamma$ is the gravitational constant, $r$ is the distance from the center 
of a star, $M(r)$ is the mass of a star part in the $0 \leq r' \leq r$ region 
of the star (the star is spherically symmetrical). 

The gravitational force is directed toward the star center, that is, this 
force has the $\rho$-projection (toward the axis of rotation). Therefore, this force 
competes with the centrifugal force (the actions of these forces are opposite) 
in the processes of separation of charges and generation of electric 
and magnetic fields. It is clear from the general reasoning, that at the given 
angular frequency and starry composition the magnetic field 
reverses its direction at some "critical" mass (or at "critical" 
$R_{0c}$); or the magnetic field of a certain star reverses its direction 
at some certain critical angular frequency $\Omega_{c}$. In the case of 
gravitational confinement, the (more light) electron cloud is distributed 
near the boundary (and just near the axis), but the (more heavy) ion layer 
is distributed near the center (the value $kT$ is the same for all particles 
and $\approx U_g$). 

The potential energy of some particle with mass $m$ is:
\begin{equation}
U_{g} = \gamma{M(r)m\over r}~,
\end{equation}
where $r = \sqrt{\rho^{2} + z^{2}}$.

In the general case the distribution of charges includes three
charged regions. The negatively charged region is placed near the axis of 
rotation and the system boundary; the positively charged region (torus) 
is placed near (and far) the radius which can be determined from the condition 
$F_g=F_c$. The bounds between the charged regions can be found 
from the equation $n_{e}(\rho,z) = \sum_{i} n_{i}(\rho,z)$. For example, 
it follows that the equation for plasma consisting of particles with one 
number $N$ is:
$$
n_{0e}\exp\Biggl ( {1\over kT}\Biggl [ \int_0^z {eH_{\rho}\over
c}\Omega\rho_1dz_1-\int_0^{\rho} {eH_z\over
c}\Omega\rho_1d\rho_1-e\int{\bf E'}_{0}d{\bf r'}+
$$
$$
{m_{e}\Omega^{2}\rho^{2}\over 2}-\gamma{M(r)m_{e}\over r}\Biggr ]\Biggr )= 
$$
$$
n_{0i}\exp\Biggl ( {1\over kT}\Biggl [
{M_{i}\Omega^{2}\rho^{2}\over 2}+Ne\int
{\bf E'}_{0}d{\bf r'}+\int_0^{\rho} {eH_z\over
c}\Omega\rho_1d\rho_1 - 
$$
\begin{equation}
\int_0^z {eH_{\rho}\over
c}\Omega\rho_1dz_1 - 
\gamma{M(r)M_{i}\over r}\Biggr ] \Biggr ) .
\end{equation}

In the general case the components ${\bf E}_{0}(\rho,z)$ can be found from the 
system of integral equations: 
\begin{equation} 
E_{0\rho}(\rho,z) = e\int_{(V)} {q(\rho_{1},
z_{1})\over R'^3}\rho_{1}[\rho-\rho_1cos(\varphi_1-\varphi)]d\rho_{1}dz_{1}d\varphi_1
~, 
\end{equation}
\begin{equation}
E_{0z}(\rho,z) = e\int_{(V)} {q(\rho_{1}, z_{1})\over R'^3}(z-z_1)\rho_{1}d\rho_{1}dz_{1}d\varphi_1~ , 
\end{equation} 
$$
R' = \sqrt{(z-z_1)^2+\rho^2+\rho_1^2-2\rho\rho_1cos(\varphi_1-\varphi)}
$$
with additional term $\gamma M(r)M_i/ r$ in
$U_i$ and $\gamma M(r)m_e/ r$ in $U_e$.

The magnetic field can be obtained from the following expressions:
\begin{equation}
H_{\rho} = {1\over c}\int_{(V)}
{j_{\varphi}(\rho_1,z_1)\over R'^3}(z-z_1)\rho_1d\rho_1dz_1d\varphi_1~ , 
\end{equation}
\begin{equation}
H_{z} = {1\over c}\int_{(V)}
{j_{\varphi}(\rho_{1},z_{1})\over
R'^3}\rho_{1}[\rho_1cos(\varphi_1-\varphi)-\rho]d\rho_{1}dz_{1}d\varphi_1
~,
\end{equation}
In the case of unsteady rotation $(\Omega^{2}\rho^{2}/ 2) \rightarrow 
\int_{0}^{\rho} \Omega (\rho',z)\rho'd\rho'$. 

As a first approximation, the effect is proportional to the particle's 
concentration (or, the ion surplus $-\Delta$). There exists such a value 
$T_m$, that $\mid H_z(T_m)\mid= max$. At first, the value $H_z$ increases 
(to a maximum $H_z(\Omega^1_m)>0$) with increasing the value $\Omega$ to 
$\Omega^1_m$; furthermore, the value $H_z$ decreases (to a minimum 
$H_z(\Omega^2_m)<0$) with increasing the value $\Omega$, and, finally, 
the value $H_z$ tends to zero with further increasing $\Omega$. We can write 
approximately
$$
{\partial H_z\over\partial T} = {H_z\over 2T} - {\Omega\over 2T}{\partial H_z
\over\partial\Omega}.
$$
The dependence $H_z(\Omega,T)$ can be represented as "a hill passing to a 
pit". Analogously, the dependence $H_z(M,T)$ can be described as "a pit 
passing to a hill". However, in the real case the value $T$ can increase with 
increasing the body mass $M$, and the value $H_z$ can be finite ($\ne 0$) 
with increasing the value $M$. 

There is another qualitative difference between the cases of star and 
rotating plasma. The star is a nonequilibrium system. At some point 
of the star the radiatmon from and to the center of the star 
are not compensated, thus causing the light pressure other than zero to exist. The 
action of this light pressure for particles depends on the effective section 
of interaction of these particles. Therefore, tle light pressure force for 
ions is larger, than that for electrons, and this influence on the charge 
separation process is opposite to the gravitational force action. 

Given the effective sections of interactions $\sigma_{e}$ and $\sigma_{i}$ and 
the light pressure function $P(\rho,z)$, the field ${\bf H}$ can be obtained in 
terms of this effect. To do this, the following substitutions need to be done: 
$$
{-m_{e}\Omega^{2}\rho^{2}\over 2} \rightarrow {-m_{e}\Omega^{2}\rho^{2}\over 2} - \sigma_{e}\int_{(0,0)}^{(\rho,z)} P(\rho',z')dr';
$$
$$
{-M_{i}\Omega^{2}\rho^{2}\over 2} \rightarrow {-M_{i}\Omega^{2}\rho^{2}\over 2} - \sigma_{i}\int_{(0.0)}^{(\rho,z)} P(\rho',z')dr'.
$$
Some important remarks will be made in section 2.4. 

\subsection{The planetary magnetism}

In this case the radiation can be ignored. In some instances the gravitational 
force can be ignored only when the of charge separation region is small, or, 
more precisely, its action is reduced to the action of pressure on the 
boundary of this region. ${\bf \Omega}$ and $N_{i}$ quantities for the planets may be 
larger, than those for the other astronomical objects. The distribution of 
electron (or ion surplus) is determined by characteristics of planetary 
system (Boltzmann factor depends on the solar mass, chemical composition etc). 
For some 
planets with strong magnetic field the radiation belts (the Earth radiation belt, 
for example, which is the rotating plasma substantially) have influence on the magnetic 
field. Note, that radiation belts can be considered as a separate system (with 
their own boundary conditions), for which all derived formulae are applicable. 

The case of planets has some qualitative difference from the other cases. The 
planets are relatively cold objects. Therefore, two remarks are in order. 
First, not all particles are involved in the processes of charge separation and of 
the magnetic field appearance. The neutral particles influence 
on the electric field ${\bf E}_{0}(\rho,z)$. To take into account this influence, the 
following substitution needs to be done in the denominators of equations (11), 
(12), (13), (14): $R'^3 \rightarrow R'^3\varepsilon (\rho,z)$, 
where $\varepsilon (\rho,z)$ is the permittivity. The polarization of 
nonmagnetic materials can influence on the magnetic field. 

Second, the influence of magnetic materials in the surface layer of a planet 
on the magnetic field at temperature lower than the Curie temperature must 
be taken into account:
$${\bf B} = \mu{\bf H} + 4\pi{\bf M}_{0},
$$
where ${\bf H}$ is the magnetic field from (13) or (14), $\mu$ is the permeability, 
${\bf M}_{0}$ is the remanent magnetization.

The latter factor can become the only one which maintains the existence of 
the magnetic field as the axial rotation of a planet is decelerated.
The qualitative picture, described above in previous subsection, is valid for 
the charge distribution and the dependence of magnetic field on different 
factors. Note, that (an agreement with this paper) there exists the electron 
surplus near the Earth's surface (see The Pysical Values, 1991). The most 
significant remarks with practical viewpoint is made in the next subsection. 

\subsection{Some remarks and estimations}

First of all we note that the Kauling theorem does not valid for models 
described. The electric field possesses the $E_{0\rho}$ and $E_{0z}$ 
components only ($E_{0\varphi}=0$). However, the nature of currents does not
determined by the magnetic force ${1\over
c}[{\bf v}\times{\bf B}]$. The existence of currents is the
consequence of the system rotation: portions of charged
particles (which form charged regions) are involved in the
rotation. In spite of particles movement (rotation) with the
same ${\bf \Omega}$ (without friction), currents can have the 
mutually reverse directions. There is a surface with
${\bf j}=0$; the field ${\bf B}$ can be equal zero on some
surface; $rot~{\bf B}$ can also be equal zero. Thus, the
principal conclusion consists in the fact, that the magnetic
field can origin in a system with cylindrical symmetry. 

Second remark needs to be made in connection with conditions
of application of models. It is easy to make an estimate of the field 
produced by the effect described for $T=0$. The magnetic field (in Gaussian 
units) is $H\sim RJ/c$, 
here $R$ is the system radius, $J\sim \rho v$. If the gravitational 
force is most important, then $eE\sim M_ig',~~\rho\sim M_ig'/(eR)$ 
here $g'\sim j\rho_bR$, where $j$ is the gravitational constant, $\rho_b$ 
is the body density. Substituting typical numbers ($M_i\sim 10^{-23}$gram, 
$v/c\sim 10^{-4}, e\sim 10^{-10}$esu, $\rho_b\sim 10^{-3}$gram/cm$^3$), 
we have $H\sim 10^{-4}$ Gauss for $R\sim 10^9$a.u. However, the model is 
statistical (principally): statistical (dynamical) equilibrium holds by 
heat movements, and the originated electric force cannot compensate 
the charge density (for example, the noncompensated force $F_g$ exists, but 
the earth atmosphere doesn't "fall" on the Earth). 
The estimations needs to be corrected for stellar and planetary magnetism. 
To do this, two ways exist. 1) The above estimations suppose that the 
magnetic field is produced by a charge surplus. Note, that 
if for an earthlike body $H\sim 1~ -~ 10^{-4}$Gauss, then the surplus 
charge density $\rho$ needs to be equal 
$\rho\sim 10^{-4} ~- ~10^{-8}$esu/cm$^3$, that is the number of charged 
particles is $10^2 ~- ~10^6$ per cm$^3$. It is possible for the earth 
phenomena (see The Physical Values, 1991).
2) The "elucidation" of the gravitational force for system neutral as a 
whole does not meant that the body will scatter under influence of electric 
force. These forces do not compensate. Besides, any liquid or rigid body 
can possess rather great portion of charges without interruption. 
The "capacitor" (described above the double layer) discharges.
To account the latter effect, one can use some potential 
(barrier maintaining a "contact potential difference") 
which counteracts the discharge process. Note as a remark, that Earth 
structure (in chemical and material composition) can be determined by 
self-consistent manner (using the chemical potential $\mu$). The presence 
of charged particles can be described as a "dilute solution", and the 
equilibrium condition for different phases is: $\mu + U = const$. 

The following remark concerns the role of electromagnetic forces for Earth 
phenomena. As a consequence of long-range (rather strong) electric 
interaction, some small charge surplus is suffice for comparability the 
electric energy with the motion energy in atmospheric (different winds) and 
hydrodynamical (sea currents) 
phenomena. To account this, one must additionally write: 1) the motion equation for 
charged particles, and, 2) the equation for the interaction of charged 
particles and neutral gas (or liquid) masses. Cosmic rays influence on the 
earth activity (earthquake, vulcan activity), since the presence of magnetic 
materials (hysteresis) leads to heat production in inner earth regions. 

\section{Magnetic fields and some factors}

\subsection{Self-sustained equilibrium rotation}

It is shown above, that in a system possessing some
fraction of free charges a partial separation of charges
occurs under an effect of centrifugal and gravitational forces
($n_e(\rho , z)$ does not coincide with $n_i(\rho , z)$) and,
as a result, the electric field ${\bf E}(\rho , z)$ arises.
Due to system's rotation the magnetic effect of currents of
these charged regions is not mutually compensated, and the
magnetic field ${\bf H}$ arises. In this case the rotation
frequency $\Omega (\rho , z)$ was assumed to be specified. 

The statement of the problem may be changed. For this purpose
we notice that the field ${\bf H}$ is directed along the axis
of rotation (axis $z$), and the field ${\bf E}$ has two
components $E_{\rho}$ and $E_z$ only, i.e. it has a component
directed towards the system's axis of rotation. As a result,
the drift of particles will take place in cross fields
$E_{0\rho}{\bf e}_{\rho}$ and $H{\bf e}_z$. This drift motion
will occur over circles around the axis of rotation. Hence, it
is worthwhile to find such a rotation rate, that resulting
electric and magnetic fields lead to a drift rotation with the
same rate, i.e. one should find stable rotational states of a
charged particles in a self-consistent manner. Note, that the (averaged) 
movement of charged particles must be close to the rotation of the 
system as a whole (a consequence of the system stability).

From condition $v_d=\Omega\rho$ and the expression for drift
velocity 
$$
{\bf v}_d = -{{\bf e}_{\varphi}cE_{0\rho}(\rho , z)\over
H(\rho , z)}
$$
the unknown rate is determined as 
\begin{equation}
\Omega = {cE_{0\rho}\over \rho [H_0(\rho , z) + H_1]} , 
\end{equation}
where $\rho$ is the distance to the axis, $c$ is the light
velocity, $H_1$ is an outer or intrinsic magnetic field ($H_1
> H_0$), and
quantities $E_{0\rho}$ and $H_0$ can be taken from previous section. 

\subsection{Dynamics of magnetic fields}

Let us imagine that the external magnetic field ${\bf H}_1$
suddenly increases: ${\bf H}_1 \rightarrow {\bf H}_1 +
{\bf H}'_1$. As a result, as follows from (15), the equilibrium
frequency of rotation of the object (or a fraction of charged
particles) decreases in this case. This decrease of
$\Omega(\rho , z)$ results in decreasing the field ${\bf H}_0
\rightarrow {\bf H}_0 - {\bf H}'_0$ and, hence, the resulting
field slightly lowers: ${\bf H} = {\bf H}_0 + {\bf H}_1 +
{\bf H}'_1 - {\bf H}'_0$. This is a consequence of the general
Le Shatelier principle. Similarly, the weakening of
the external magnetic field ${\bf H}_1$ leads to increasing
the drift rotation rate, which, in its turn, causes the
increase of the field ${\bf H}_0$, i.e. the processes
weakening the external action occur again (the system tends to
conserve its equilibrium state). 

Note that at equilibrium either 
$\Omega = const$, or $\Omega = \Omega (\rho , z)$: 
\begin{equation}
{\Omega^2\rho_1^2\over 2} \rightarrow \int_0^\rho 
\Omega(\rho',z)\rho'd\rho', 
\end{equation}
for charged particles and the source of energy and the mechanism of compensation of heat 
losses (for friction) must exist. If the state of a system differs 
from equilibrium one by any reason (for example, the external 
forces are acting, or some processes occur inside the object, which lead to 
redistribution of charges), then, in principle, one can find the forces, 
tending to bring the system into equilibrium, and describe the dynamics 
(of fields, of rotation rate). 

It follows from (15), that as the rotation frequency of an astronomical object 
changes adiabatically, the electric and magnetic fields also change: 
$$
\dot{\Omega} = {c\over \rho}[{\dot{E}_{0\rho}\over H} - {E_{0\rho}\dot{H}\over H^2}]. 
$$
As the first approximation, using 
$$
\dot{E}_{0\rho}={\partial E_{0\rho}\over\partial\Omega}\dot{\Omega}+
{\partial E_{0\rho}\over\partial\Delta}\dot{\Delta}+{\partial E_{0\rho}\over
\partial T}\dot{T} ,~~~~~~ {\partial E_{0\rho}\over\partial\Omega}=-{2T\over\Omega}
{\partial E_{0\rho}\over\partial T} ,
$$
we have 
\begin{equation}
\dot{H} = H\Biggl ( {\dot{\Delta}\over\Delta}-{\dot{\Omega}\over\Omega}\Biggr ) 
+{H\over E_{0\rho}}{\partial E_{0\rho}\over \partial T}\Biggl ( \dot{T}-2T
{\dot{\Omega}\over\Omega}\Biggr ) . 
\end{equation}
The value ${\partial E_{0\rho}\over\partial T}$ must be evaluated from the 
self-consistent system of equations (from Sec.2). Since $H=H_0+H_1$, the 
hysteresis influences on the field dynamics. The change of the magnetic field 
polarity may occur (without changing the direction of astronomical object's 
rotation), when 
the direction of the electric field $E_{0\rho}$ inside the system changes. 
The change of polarity for a system with magnetic materials (there exists 
some threshold due to hysteresis) requires long-lasting influence (cosmic 
rays can change the total charge of the system, for example). 

In principle, one can obtain the solution to the problem of fields dynamics, 
in the case of "sufficiently slow" variation of various external and internal 
factors. It is apparent that the dynamics of 
$T, \rho , (V)$ and $\Omega$ ~determines the dynamics of electric and magnetic fields 
by formulae (15) - (17). Therefore, to finally determine the fields dynamics 
and the rotation rate, one should also add the following 
equations:\\
1. the moment of external forces acting on a system determines the additional 
change of rotation frequency $\Omega$ (if the object does not possess central 
symmetry); this factor may cause the magnetic field precession; \\
2. the law of variation of object's form $(V)$ (this factor is insignificant 
for stars and planets); \\
3. the law of variation of density $\rho ({\bf R}, t)$; \\
4. the law determining the temperature variation $T({\bf R}, t)$;\\ 
5. the law of (external) change of the total charge (this factor is the 
most important one). 

Now it is clear that the characteristic times of variation of all these 
quantities must be higher than characteristic times of diffusion for a 
fraction of charged particles participating in the fields generation. In 
the opposite case one must take into account diffusion terms in equations 
for $\rho , T, \Omega$ and $(V)$ (i.e. the equation of diffusion should be 
written separately for a fraction of charged particles). In this case 
the hysteresis phenomena are characteristic of the fields variation. 

How the object driven from the state of equilibrium will behave? 
From general considerations, the transition into a new equilibrium state (or 
into in old one, if the factor acts for a limited time) may occur either 
asymptotically, or in an oscillational manner (with polarity change, in 
particular), depending on the fact, whether the characteristic time of 
system oscillation ($\Omega^{-1}$) is higher (or, respectively, considerably 
lower) than that of system's energy dissipation. 

In the case, if the rotation frequency inside an astronomical object is lower 
than the equilibrium frequency possible for the given system, then the new 
equilibrium state will correspond to the absence of rotation. Because of the 
dependence $\Omega (\rho , z) \ne const$ (caused by the fields) the friction 
will arise between different sections of the object and, as a result of energy 
dissipation, the rotation of an astronomical object will be slowed down. 
This is just an internal mechanism of slowing down the rotation of 
astronomical objects (and warming-up of inner regions). 

\section{Conclusions}

The general statistical approach to the magnetism problem is
introduced. It can help to obtain the value of the magnetic field to 
sufficient accuracy (from the exact integral solution). 
The mechanism of the magnetic field generation in astronomical objects 
(nebulae, stars, planets) can be explained in the following way: in the 
general case the centrifugal force, the gravitational force and the light pressure 
act on dissimilar particles in a different manner, causing different $\rho$-distributions 
of these particles and giving rise to the partial separation of charges; the electric field 
can be found from formulae (11), (12) (see the substitutions in the 
text); the rotation of these objects produces (from these charges) circulating 
currents, the magnetic action of which is not compensated, and the magnetic 
field can be found from formulae (13), (14) (see the 
substitutions in the text). Generally speaking, the existence of the magnetic field 
dispenses with the need for conductive regions; the availability of some 
portion of free charges would be sufficient for this mechanism. 

The equilibrium frequency of rotation of an astronomical object (see (15)) 
may be found in a self-consistent manner. Knowing the factors, 
which influence the density, 
temperature, rotation of an object and external fields, one can determine in 
this case the dynamics of system's magnetic field. 

The further applications of formulae obtained might be associated with numerical 
methods or with some simplifications which take into account some specific data.

{}

\end{document}